\documentclass[iop]{emulateapj}
\usepackage{amsmath}
\usepackage{natbib}
\usepackage{textcomp}
\usepackage{color}
\usepackage[colorlinks=true,citecolor=blue]{hyperref}
\usepackage{graphicx,color}
\usepackage{times}
\usepackage{amssymb}

\newcommand{\lum}{erg\,s$^{-1}$}
\newcommand{\phflux}{\mbox{${\rm \, ph \,\, cm^{-2} \, s^{-1}}$}}
\newcommand{\ergflux}{\mbox{${\rm \, erg \,\, cm^{-2} \, s^{-1}}$}}
\newcommand{\fermi}{{\it Fermi}}
\newcommand{\swift}{{\it Swift}}

\newcommand{\nlsy}{$\gamma$-NLSy1}
\newcommand{\gm}{$\gamma$}

\slugcomment{ApJ accepted}

\shorttitle{SBS 0846+513}
\shortauthors{Paliya et al.}

\begin{document}

\title{Broadband Observations of the Gamma-ray Emitting Narrow Line Seyfert 1 Galaxy SBS 0846+513}

\author{Vaidehi S. Paliya$^{1,\,2}$, Bhoomika Rajput$^1$, C. S. Stalin$^1$, and S. B. Pandey$^3$}
\affil{$^1$Indian Institute of Astrophysics, Block II, Koramangala, Bangalore-560034, India}
\affil{$^2$Department of Physics, University of Calicut, Malappuram-673635, India}
\affil{$^3$Aryabhatta Research Institute of Observational Sciences, Manora peak, Nainital-263129, India}
\email{vaidehi@iiap.res.in}

\begin{abstract}
We present the results of our broadband study of the \gm-ray emitting narrow line Seyfert 1 (NLSy1) galaxy SBS 0846+513 ($z=0.585$). This includes multi-band flux variations, \gm-ray spectral analysis, broad band spectral energy distribution (SED) modeling, and intranight optical variability (INOV) observations carried over 6 nights between 2012 November and 2013 March using the 2 m Himalayan Chandra Telescope and the 1.3 m telescope at Devasthal. Multiple episodes of flaring activity are seen in the \gm-ray light curve of the source which are also reflected in the observations at lower frequencies. A statistically significant curvature is noticed in the seven years averaged \gm-ray spectrum, thus indicating its similarity with powerful flat spectrum radio quasars (FSRQs). Modeling the SEDs with a one zone leptonic emission model hints the optical-UV spectrum to be dominated by synchrotron radiation, whereas, inverse Compton scattering of broad line region photons reproduces the \gm-ray part of the SEDs. The source was found to be variable on all the six nights of optical observations with a variation of $\sim$0.3 magnitude within a single night, coinciding with a high \gm-ray activity state. The observed large amplitude INOV clearly indicates the presence of a closely aligned beamed relativistic jet in SBS 0846+513. Our broadband study supports the recent claims in literature that \gm-ray emitting NLSy1 galaxies are similar to blazars and constitute the low black hole mass counterparts to FSRQs.
\end{abstract}


\section{Introduction}\label{sec1}
The high energy extragalactic \gm-ray sky, as seen by the \fermi~Gamma-ray Space Telescope,
 is dominated by the blazar class of active galactic nuclei \citep[AGN, e.g.,][]{2015ApJ...810...14A}. Along with blazars, significant \gm-ray emission is  also observed from about half-a-dozen radio-loud narrow 
line Seyfert 1 (RL-NLSy1) galaxies \citep[e.g.,][]{2009ApJ...707L.142A,2011nlsg.confE..24F,2012MNRAS.426..317D,2015MNRAS.454L..16Y}.
 NLSy1 galaxies are a sub-class of AGN with peculiar properties that distinguishes them from the conventional 
broad line Seyfert 1 galaxies. For example, their optical spectra consist of 
narrow Balmer lines (FWHM (H$_{\beta}$) $<$ 2000 km s$^{-1}$), weak 
[O~{\sc iii}] ([O~{\sc iii}]/H$_{\beta} <$ 3), and strong optical 
Fe~{\sc ii} lines \citep[][]{1985ApJ...297..166O,1989ApJ...342..224G}. They 
exhibit rapid X-ray flux variations \citep[][]{1995MNRAS.277L...5P} and steep soft 
X-ray spectra \citep[][]{1996A&A...305...53B}. It is found that NLSy1 galaxies, in 
general, host low-mass black holes ($\sim 10^{6}-10^{8} M_{\odot}$) accreting 
close to the Eddington limit \citep[][]{2004ApJ...606L..41G,2012AJ....143...83X}. 
About 7\% of the NLSy1 galaxy population are found to be radio-loud and many of 
them exhibit compact core-jet radio morphology, flat/inverted radio spectra, 
high brightness temperature, and superluminal patterns \citep[][]{2006AJ....132..531K,2006PASJ...58..829D}. These features hint for the presence of closely aligned relativistic 
jets in them and observations from the \fermi-Large Area Telescope 
(\fermi-LAT) provided the confirmation. Therefore, it is of great interest to 
test their similarities/dissimilarities with the blazar class of AGN, which 
emits copiously in the \gm-ray band. The 
optical and infrared (IR) flux variations of some of these \gm-ray emitting 
NLSy1 (\nlsy) galaxies are found to be similar to blazars \citep[][]{2010ApJ...715L.113L,2012ApJ...759L..31J,2013MNRAS.428.2450P}. Variable optical polarization, similar to 
that known in several blazars is also observed in some \gm-ray emitting NLSy1 
galaxies \citep[][]{2013ApJ...775L..26I,2014PASJ...66..108I}. 
Blazars are 
known to show hr-scale \gm-ray flux variations \citep[e.g.,][]{2011A&A...530A..77F}
and a similar short timescale \gm-ray variability was noticed from a NLSy1 
galaxy 1H 0323+342 ($z=0.063$) during its 2013 September outburst \citep[][]{2014ApJ...789..143P}. A statistically significant curvature is observed in the \gm-ray 
spectra of few \nlsy~galaxies \citep[][]{2015AJ....149...41P} which indicates their 
resemblance more with the FSRQ class of blazars. The broad-band 
spectral energy distributions (SEDs) of \nlsy~galaxies exhibit the typical 
double hump structure \citep[e.g.,][]{2009ApJ...707L.142A,2012MNRAS.426..317D,2013ApJ...768...52P}, similar to blazars. Additionally, the requirement of external 
Compton (EC) mechanism to explain the \gm-ray window of the SED provides 
another supporting evidence of the resemblance of these objects with FSRQs 
\citep[e.g.,][]{2009ApJ...707L.142A,2012MNRAS.426..317D,2012A&A...548A.106F,2013ApJ...768...52P}. 

SBS 0846+513 (R.A.=08:49:58.0, J2000; Decl.=+51:08:28, J2000; $z=0.585$) is a 
NLSy1 galaxy found to be a \gm-ray emitter by \citet{2011nlsg.confE..24F}. \fermi-LAT has 
observed more than one episodes of \gm-ray flaring activity from this object 
where the isotropic \gm-ray luminosity reaches as high as 10$^{48}$ 
\lum~\citep[e.g.,][]{2012MNRAS.426..317D}. \citet{2014ApJ...794...93M} have reported  
a high amplitude three-magnitude optical flare in 2013 April, closely 
associated with a high \gm-ray state. They also reported the detection of 
high optical polarization ($\sim$10\%) during the same period. The 
very high brightness temperature \citep[$T_{\rm b}>$10$^{13}$ K;][]{2005ChJAA...5...41Z}, 
large amplitude optical variability, and the detection of superluminal motion 
\citep[][]{2012MNRAS.426..317D} further hint for the presence of closely aligned 
relativistic jet in this source. In this work, using all the publicly 
available multi-wavelength data taken during the first seven years of 
\fermi~operation, we study this source in detail. New observations
of its intra-night optical variability (INOV) characteristics are also presented. Considering a leptonic origin of radiation, we present a 
consistent picture of the physical properties of this source.

The paper is organized as follows. We describe the adopted data reduction 
procedure in section~\ref{sec_data}. The obtained results are presented and discussed 
in section~\ref{sec_results} followed by the conclusion in section~\ref{sec_con}. We use 
the cosmological parameters $\Omega_{m}$=0.27 and Hubble constant 
{\it H}$_{0}$ = 71 km s$^{-1}$ Mpc$^{-1}$. 

\section{Multi-wavelength Observations and Data Reductions}\label{sec_data}
\subsection{\fermi-Large Area Telescope Observations}
We analyze LAT observations of SBS 0846+513 covering the first seven years of 
\fermi~monitoring, i.e., MJD 54683$-$57222. We follow the standard data 
reduction procedure as described in the online documentation\footnote{http:
//fermi.gsfc.nasa.gov/ssc/data/analysis/documentation/} and here it is briefly mentioned. We use the latest data set (Pass 8) and, in the energy 
range of 0.1$-$300 GeV, events belonging to {\tt evclass} 128 and {\tt evtype} 
3 and lying within 15$^{\circ}$ source region are considered. Data analysis is 
performed with the package {\tt ScienceTools v10r0p5} along with the use of 
instrument response functions P8R2\_SOURCE\_V6. To avoid the contamination 
from Earth limb \gm-rays, the maximum zenith angle is adopted as 90$^{\circ}$. The 
region of interest (ROI) is defined as a circle of 10$^{\circ}$ radius 
centered at the \gm-ray position of SBS 0846+513, as defined in the third 
catalog of \fermi-LAT detected objects \citep[3FGL;][]{2015ApJS..218...23A}. A maximum 
likelihood test statistics (TS) = 2$\Delta$log($\mathcal{L}$), where 
$\mathcal{L}$ represents the likelihood function between models with and 
without a point source at the source position, is derived to determine the 
significance of the \gm-ray signal. All the sources present within ROI 
and defined in the 3FGL catalog are considered for the spectral analysis with 
their parameters left free to vary during the fitting. In addition to that, 
objects lying between 10$^{\circ}$ to 15$^{\circ}$ from the source of interest,
are also included, however, their parameters are kept fixed to the 3FGL 
catalog values. Galactic and the isotropic extra-galactic background emissions
are also properly accounted using the latest background templates 
({\tt gll\_iem\_v06.fit} and {iso\_P8R2\_SOURCE\_V6\_v06.txt}, respectively). 
A binned likelihood analysis is performed over the period of interest and all 
the sources with TS$<$25 are removed. This updated model is then used for 
further temporal and spectral studies.

All the errors associated with the LAT data analysis are 1$\sigma$ statistical 
uncertainties, unless specified. Primarily governed by the uncertainty in the 
effective area, the systematic uncertainties on the measured fluxes are energy 
dependent: it amounts to 10\% at 30 MeV, decreasing linearly as a function of 
log(E) to 3\% at 100 MeV, and increases linearly as a function of log(E) 
beyond 100 GeV to about 15\% at 1 
TeV\footnote{http://fermi.gsfc.nasa.gov/ssc/data/analysis/LAT\_caveats.html}.

\subsection{\swift~Telescope Observations}
The \swift~telescope \citep[][]{2004ApJ...611.1005G} has observed SBS 0846+513 for a 
total of 14 times in the first seven years of \fermi~mission (covering MJD 
54683$-$57222 or 2008 August 4 to 2015 July 15). The source was observed using 
all the three instruments on board: the Burst Alert Telescope \citep[BAT;][15$-$150 keV]{2005SSRv..120..143B}, the X-Ray Telescope \citep[XRT;][0.3$-$10 keV]{2005SSRv..120..165B}, and the Ultraviolet/Optical Telescope \citep[UVOT;][170$-$600 nm]{2005SSRv..120...95R}.

SBS 0846+513 is never detected by BAT, primarily due to poor sensitivity of 
the instrument for short exposures and also due to the low level of hard X-ray 
emission from the source. Consequently, it is not included in the 70-month \swift-BAT
hard X-ray catalog \citep[][]{2013ApJS..207...19B}.

We reduce the XRT data following standard procedures available within the 
HEASOFT package (6.17) and using the latest CALDB files. In particular, the 
tool {\tt xrtpipeline(v.0.13.1)} is used to perform standard screening and 
filtering criteria on the data set. All XRT observations were taken in the 
most sensitive photon counting mode and we use the standard grade selection of 
0$-$12. The calibrated and cleaned event files are summed to generate energy 
spectra. The source spectra are extracted from a circular region centered at 
the source and having a radii of 47$^{\prime\prime}$, whereas, the background 
events are selected from an annular region centered at the source position and 
having inner and outer radii as 55$^{\prime\prime}$ and 110$^{\prime\prime}$, 
respectively. We combined the exposure maps using the tool XIMAGE and 
ancillary response files are created with {\tt xrtmkarf} to account for 
vignetting and PSF corrections. Due to low photon statistics, to perform 
spectral analysis, we rebinned the spectra to ensure a minimum of 1 count per 
bin and use Cash statistics \citep[][]{1979ApJ...228..939C}. An absorbed power law \citep[neutral 
hydrogen column density $N_{\rm H}=2.91 \times 10^{20}$ cm$^{-2}$;][]{2005A&A...440..775K} model is used to perform the fitting within XSPEC \citep[][]{1996ASPC..101...17A}.

The observations from \swift-UVOT are integrated using the task 
{\tt uvotimsum}. Source counts are extracted from a 5$^{\prime\prime}$ 
circular region centered at the target and background is extracted from a 
larger area free from source contamination. The source magnitudes are 
extracted with the task {\tt uvotsource}. Observed magnitudes are then 
de-reddened for galactic extinction \citep[][]{2011ApJ...737..103S}. The 
conversion from corrected magnitudes to flux units is done using the zero 
points and conversion factors of \citet[][]{2011AIPC.1358..373B}.

\subsection{Optical monitoring Observations}
The \nlsy~galaxy SBS 0846+513 was observed on a total of six nights between 
2012 November and 2013 March as part of our ongoing campaign on NLSy1 
galaxies. Three nights of observations were carried out on the 1.3 m 
telescope located at Devasthal and operated by Aryabhatta Research Institute of 
Observational Sciences (ARIES), India. Another three nights of monitoring 
was done using the 2 m Himalayan Chandra Telescope at 
the Indian Astronomical Observatory, located at Hanle, India. All the 
observations were taken in the $R$ band. We follow the standard procedures in 
{\tt IRAF}\footnote{{\tt IRAF} is distributed by the National Optical Astronomy
Observatories, which is operated by the Association of Universities for 
Research in Astronomy, Inc. under co-operative agreement with the National 
Science Foundation.} to perform pre-processing of the images (e.g., bias 
subtraction, flat-fielding, and cosmic ray removal). The instrumental magnitudes
of the target and comparison stars are derived by 
aperture photometry, using the {\tt phot} task available within the APPHOT 
package in {\tt IRAF}. Following the procedures outlined in \citet{2004JApA...25....1S}, we derived the optimum aperture radius for photometric analysis. First, we generate star$-$star differential light curves for a series of aperture 
radii beginning from the median seeing FWHM of that night. The aperture that 
corresponds to the minimum scatter for the different pairs of comparison stars 
is selected as the optimum aperture. The 
positions and apparent magnitudes of the comparison stars (taken from 
USNO\footnote[2]{http://www.nofs.navy.mil/data/fchpix/}) used in the 
differential photometry, are provided in Table~\ref{tab_stars}. It should be noted that 
uncertainty in the magnitudes of the comparison stars, taken from this catalog, may be up to $\sim$0.25 
mag \citep[][]{2003AJ....125..984M}.

\section{Results and Discussion}\label{sec_results}
\subsection{Average \gm-ray spectra}
The seven years averaged \fermi-LAT data of SBS 0846+513 is fitted with both 
power law (PL) and log parabola (LP) models and the obtained results are 
presented in Table~\ref{tab:gamma}. As can be seen, the fitting of the PL model indicates a relatively harder $\gamma$-ray spectrum of SBS 0846+513 ($\Gamma_{0.1-300~{\rm GeV}}=2.23\pm0.03$) compared to other $\gamma$-NLSy1 galaxies \citep[][]{2015AJ....149...41P}. The $\sim$7 years average $\gamma$-ray flux is (3.79 $\pm$ 0.16) $\times$ 10$^{-8}$ \phflux~and this corresponds to an isotropic $\gamma$-ray luminosity ($L_{\rm \gamma}$) of $\sim$4 $\times$ 10$^{46}$ \lum. Such a high $L_{\rm \gamma}$ is typically observed from FSRQs \citep[e.g.,][]{2015ApJ...810...14A}. The associated significance of detection is $\sim$50$\sigma$ (TS $\approx$2635). Moreover, fitting of the LP model hints for the presence of curvature in the $\gamma$-ray spectrum ($\beta=0.11\pm0.02$, TS $\approx$ 2650). In general, powerful FSRQs exhibit a curved \gm-ray 
spectrum \citep[e.g.,][]{2015ApJS..218...23A} and this indicates the similarity of 
SBS 0846+513 with FSRQs.

\subsection{Multi-wavelength temporal properties}
\begin{figure}
\centerline{\includegraphics[width=14.0cm]{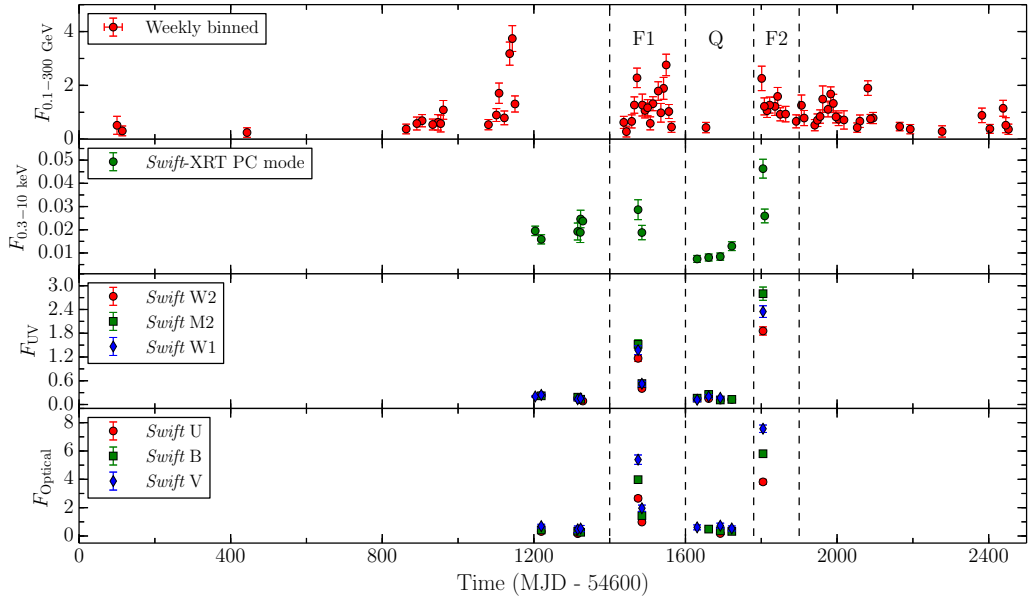}}
\caption{Multi-wavelength light curves of SBS 0846+513 covering the period 2008 August to 2015 July. Top panel represents the weekly binned \gm-ray light curve with flux units of 10$^{-7}$ \phflux. Remaining three panels correspond to \swift~observations binned as one point per observation Id. \swift-XRT data points are in units of counts s$^{-1}$ and \swift-UVOT observations are in units of 10$^{-12}$ \ergflux. F1 and F2 are two high activity periods and Q represents a low activity state, selected for SED modeling. See the text for details.\label{fig_mw_lc}} 
\end{figure}

The long term multi-wavelength temporal behavior of SBS 0846+513 is presented 
in Figure~\ref{fig_mw_lc}. In this plot, the \fermi-LAT data points are weekly binned 
and \swift~observations are shown as one point per observation Id. Multiple 
episodes of \gm-ray flares are visible and it appears from the available 
observations that the object brightened at other frequencies also during the 
period of high \gm-ray activity, though it could not be statistically 
established due to sparseness of the data. The details of these \gm-ray flaring 
behaviors are already described in \citet[][]{2015AJ....149...41P}  and 
here we concentrate more on the multi-frequency properties of SBS 0846+513. 
From the multi-wavelength light curves, we select two high activity periods 
(F1 and F2) and a quiescent period (Q) for further analysis and SED 
modeling. The choice of selecting the periods is driven by the availability of 
multi-frequency data. As can be seen in Figure~\ref{fig_mw_lc},  observations at lower 
frequencies are not available during the first flare of SBS 0846+513 (before MJD 55800), 
and thus, this flaring period is not considered for SED analysis. On the other 
hand, flares during the periods F1 and F2 are simultaneously observed by \swift, 
thus providing an unique opportunity to study them in detail. Interestingly, 
visual inspection of Figure~\ref{fig_mw_lc} suggests that the amplitude of \gm-ray 
variability during the periods F1 and F2 seems similar, however, the source 
appeared brighter during the later period in low frequency \swift~observations.

\subsection{Spectral energy distribution}

We generate SEDs for all three periods under consideration (Figure~\ref{fig_mw_lc}) by 
averaging the fluxes over each time intervals. In Table~\ref{tab:sed_flux}, we present the 
associated flux values obtained from this analysis. We model the SEDs by 
adopting an one zone leptonic emission model described in \citet{2009MNRAS.397..985G} and \citet{2009ApJ...692...32D} and here we describe it in brief. 
The emission region is assumed to be spherical, moving relativistically with 
bulk Lorentz factor $\Gamma$ at a distance $R_{\rm diss}$ from the central 
black hole of mass $M_{\rm BH}$. The semi opening angle of the conical jet is 
considered as 0.1 rad and the size of the emission region is constrained by 
assuming it to cover the entire cross-section of the jet. The emission region 
is filled with the electrons following a smoothly joining broken power law of 
the form
\begin{equation}
N'(\gamma')  \, = \, N'_0\, { (\gamma'_{\rm brk})^{-p} \over
(\gamma'/\gamma'_{\rm brk})^{p} + (\gamma'/\gamma'_{\rm brk})^{q}}
\end{equation}
where $p$ and $q$ are the energy indices before and after the break 
energy ($\gamma'_{\rm brk}$) respectively (primed quantities are measured in 
the comoving frame). The accretion disk radiation is modeled by assuming a standard optically thick, geometrically thin \citet{1973A&A....24..337S} disk. Locally, its spectrum is assumed as that emitted by a multi-temperature annular blackbody. 
The broad line region (BLR) 
is considered as a spherical shell reprocessing 10\% of the disk luminosity 
and emits like a blackbody peaking at the rest-frame Lyman-$\alpha$ 
frequency \citep[][]{2008MNRAS.386..945T}. The size of the BLR is assumed to 
follow the relation $R_{\rm BLR} = 10^{17} L^{1/2}_{\rm disk,45}$ cm, where 
$L_{\rm disk,45}$ is the accretion disk luminosity in units of $10^{45}$ 
\lum~\citep[][]{2009MNRAS.397..985G}. In our model, the presence of the 
dusty torus is also considered which reprocesses 50\% of the accretion disk 
emission at IR frequencies. For simplicity, it is assumed to be a thin 
spherical shell and its emission is considered as a simple blackbody peaking 
at the temperature $T_{\rm IR}$. The size of the torus, similar to BLR, scales 
with the square root of the disk luminosity and assumed to follow the relation 
$R_{\rm IR} = 10^{18} L^{1/2}_{\rm d,45}$ cm. In the comoving frame, the 
radiation energy densities of all the components are computed at the distance 
$R_{\rm diss}$ from the central black hole following \citet{2009MNRAS.397..985G}. The electrons radiate via synchrotron and inverse Compton 
scattering processes in the presence of a tangled but uniform magnetic field. 
The synchrotron and synchrotron self Compton (SSC) radiations are derived in 
the observer's frame, using the prescriptions of \citet{2008ApJ...686..181F}. Along 
with SSC, photons from external components such as the accretion disk, the 
BLR, and the torus also participate in the IC scattering \citep[e.g.,][]{2009ApJ...692...32D,2009MNRAS.397..985G}. The jet powers are calculated following 
\citet{2008MNRAS.385..283C}. Protons are assumed to carry only the inertia of 
the jet and do not contribute to the observed radiation. Furthermore, it should be noted that based on H$_{\beta}$ line parameters, \citet{2011ApJS..194...45S} estimated the black hole mass of SBS 0846+513 as 
$\sim$9.77 $\times$ 10$^7$ $M_{\odot}$. The broadband SED of this 
object does not show any evidence for the presence of the big blue bump (a 
characteristic signature of the accretion disk emission), so the observed 
optical spectrum probably has significant contamination from jet emission not 
taken into account in \citet{2011ApJS..194...45S}, and therefore these estimates should 
be taken with caution. Recently \citet{2015A&A...575A..13F} have performed a detailed optical spectroscopic analysis of a sample of NLSy1 galaxies, including SBS 0846+513. They used the $H_{\beta}$ line luminosity, instead of continuum at 5100 \AA, to derive $M_{\rm BH}$ and thus were able to avoid the flux contamination from the host galaxy and the jet. They found $M_{\rm BH}$ of SBS 0846+513 as 3.2 $\times$ 10$^{7}$ $M_{\odot}$. We reproduce the observed optical-UV part of the SED with a combination of the accretion disk model and synchrotron emission in such a manner so as not to overproduce the observations. With this, we derive a black hole mass as  $M_{\rm BH}=$5.5 $\times$ 10$^7$ $M_{\odot}$ and the disk luminosity as $L_{\rm disk}=$1 $\times$ 10$^{44}$ \lum. Our $M_{\rm BH}$ estimation, via SED modeling, agrees within a factor of two to that derived by \citet{2015A&A...575A..13F} and therefore, is reliable.

\begin{figure}
\hbox{
\includegraphics[width=7.0cm]{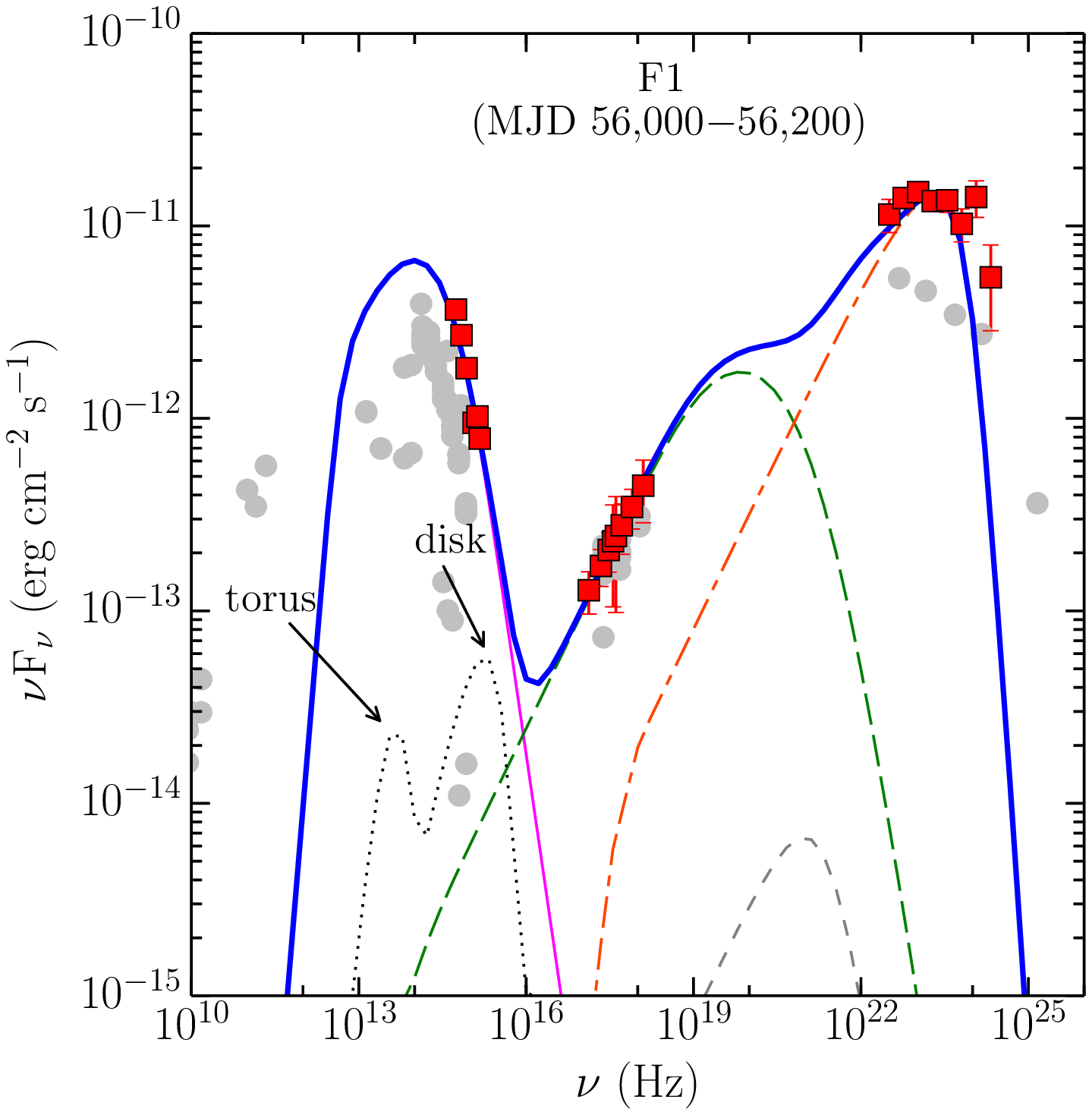}
\includegraphics[width=7.0cm]{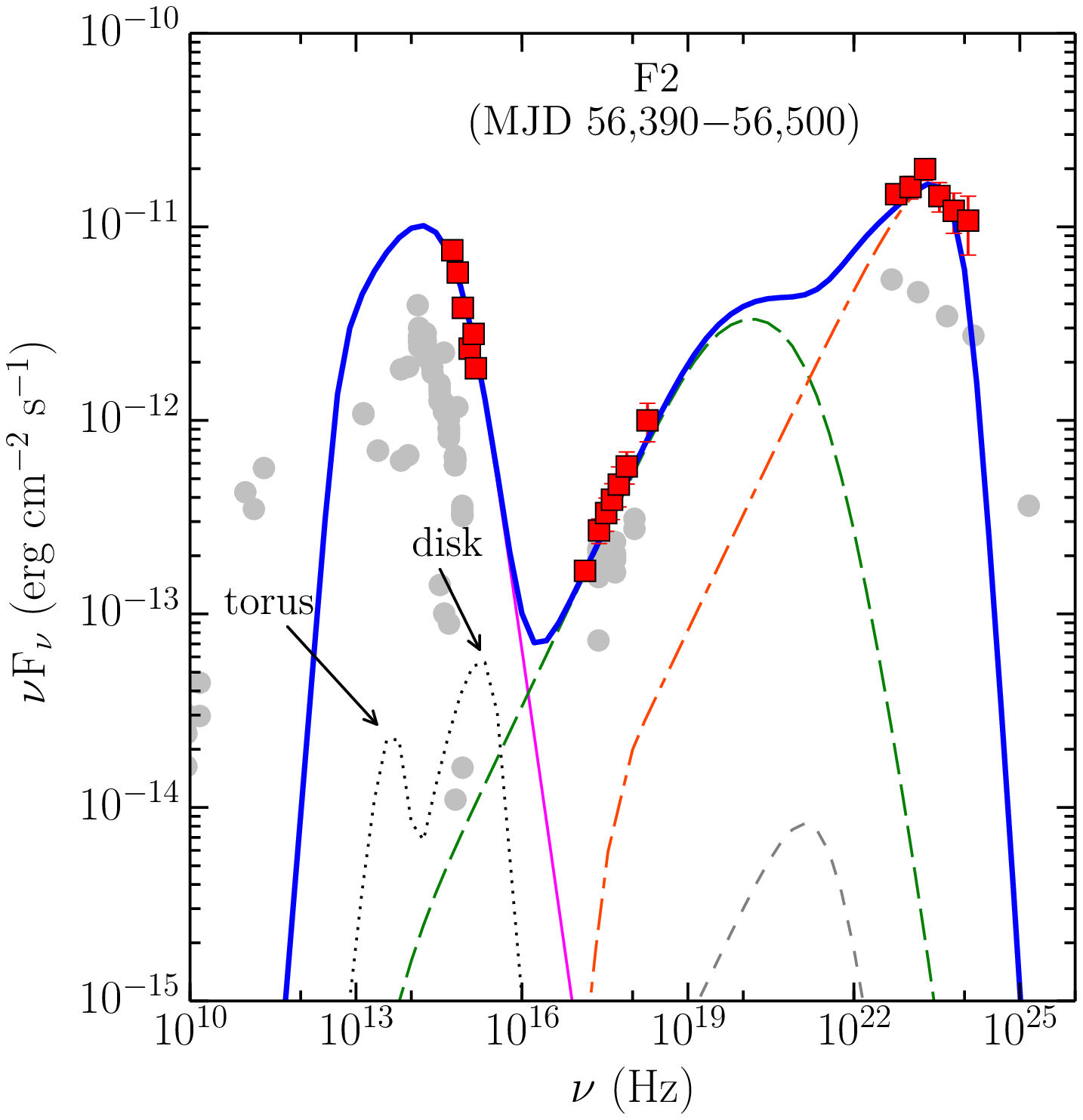}
}
\hbox{
\centerline{\includegraphics[width=7.0cm]{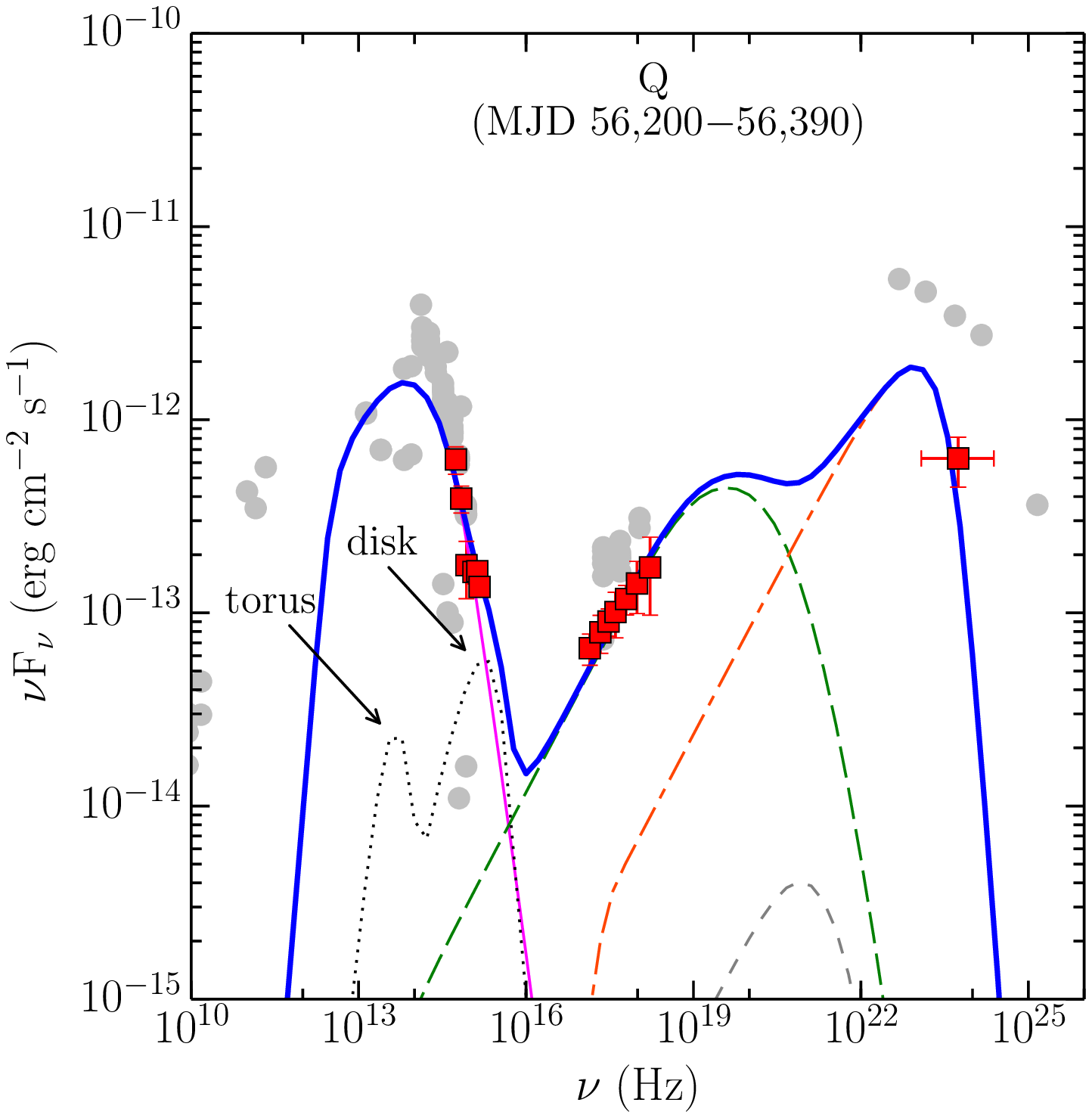}}
}
\caption{SEDs of SBS 0846+513 during various activity states. Red squares are the simultaneous observations from \swift~and \fermi-LAT, whereas, grey circles represent the archival observations. Thermal radiations from the accretion disk and the torus are shown by dotted black line. Pink thin solid, green long dashed, orange dash-dash-dot, and grey dashed lines correspond to synchrotron, SSC, EC-BLR, and EC-torus emissions, respectively. Blue thick solid line is sum of all the radiative components.\label{fig:sed}} 
\end{figure}

The modeled SEDs are shown in Figure~\ref{fig:sed} and the associated modeling parameters 
are given in Table~\ref{tab_sed_param}. In Figure~\ref{fig:sed}, the red squares represent the simultaneous 
observations from \swift~and \fermi-LAT and grey circles correspond to 
archival data\footnote{http://tools.asdc.asi.it/SED/}.

As can be seen in Figure~\ref{fig:sed}, the SED of SBS 0846+513 has a typical double hump 
structure, similar to blazars. The synchrotron radiation dominates the 
observed optical-UV spectra in all the activity states and the contribution 
from the accretion disk is negligible. This is supported by
the detection of a high optical
polarization ($>$10\%) during the 2013 April flare (associated
here with F2 state) of SBS 0846+513 \citep{2014ApJ...794...93M}. Also, the observed 
optical-UV spectrum is extremely steep which is difficult to explain by 
accretion disk model. A high synchrotron emission indicates a high level of 
SSC emission which is evident in Figure~\ref{fig:sed}. A flat X-ray spectrum is observed 
which is successfully explained by SSC mechanism. We observed a relatively hard \gm-ray spectrum in both the flaring states 
(see Table~\ref{tab:sed_flux}) which, according to our model, is reproduced by the EC-BLR 
process. Visual inspection suggests for the possible existence of a break in 
the \gm-ray spectrum which we interpret as a consequence of EC-BLR peak to 
lie within the LAT energy range. From SED modeling of both the flaring periods, we constrain the location 
of the emission region to be at the outer edge of the BLR where a major 
contribution of the seed photons for IC scattering is provided by BLR clouds. 
These results are different from that obtained by \citet{2012MNRAS.426..317D,2013MNRAS.436..191D} where the \gm-ray emission is explained on the basis of 
EC-torus mechanism and the location of the emission region was constrained 
to be far out from the BLR. Since the emission region is located relatively 
closer in our case, the derived  magnetic fields are higher and the size of 
the emission regions are smaller compared to them. 
On the other hand, during the low activity state we found the emission region to be located within the BLR.

\subsection{Intra-night optical variability}
\begin{figure}
\hbox{
\includegraphics[width=5.5cm]{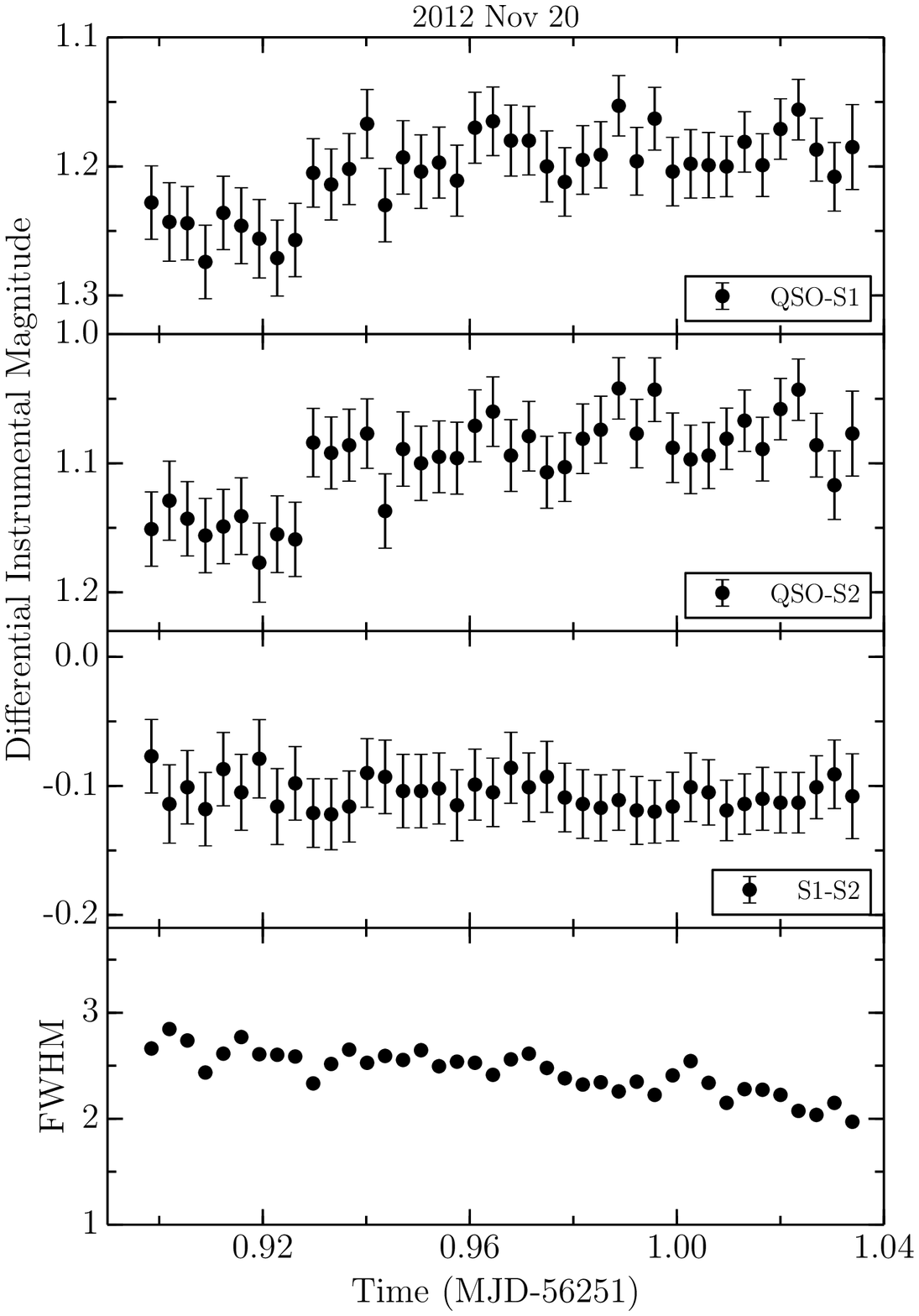}
\includegraphics[width=5.5cm]{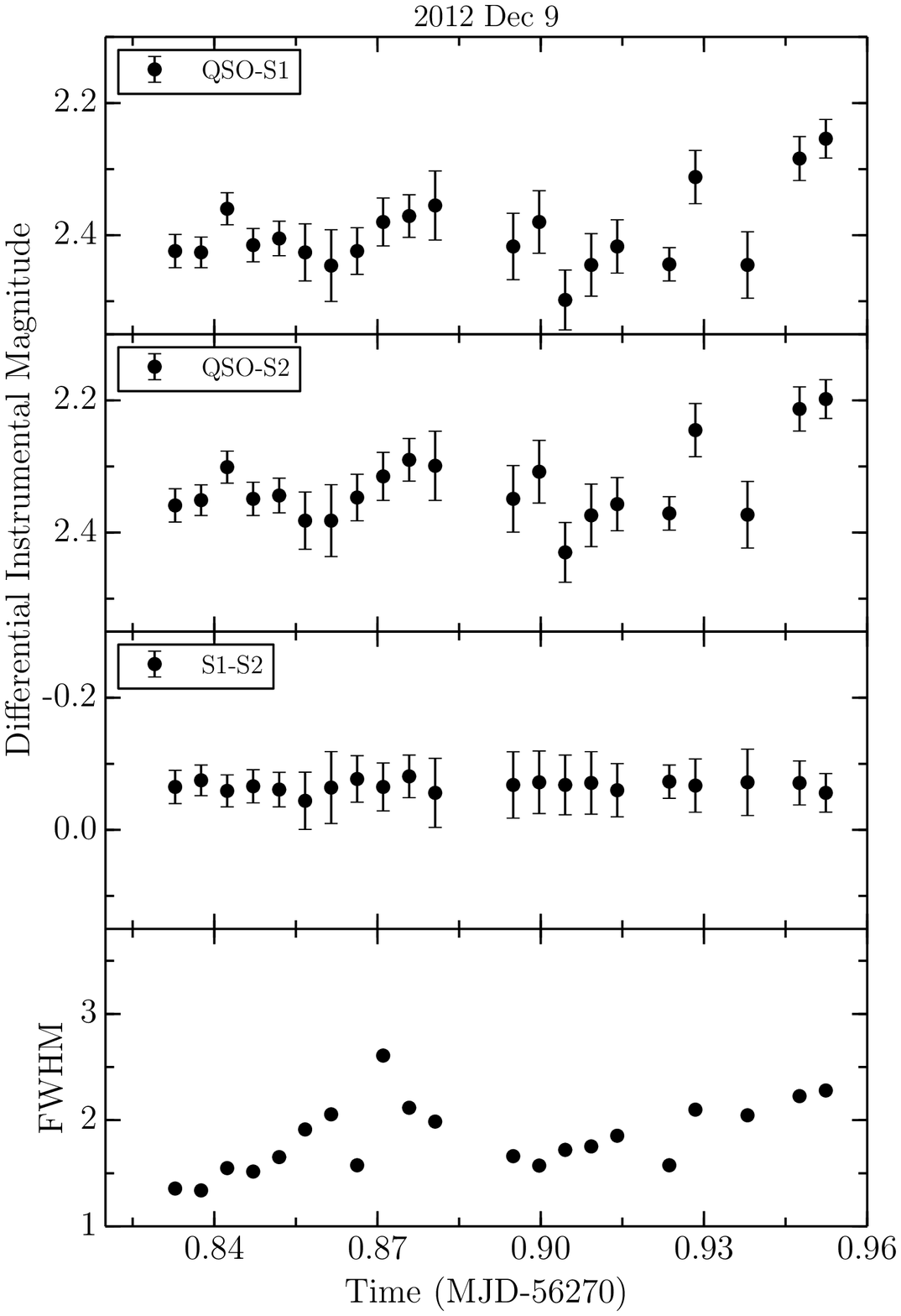}
\includegraphics[width=5.5cm]{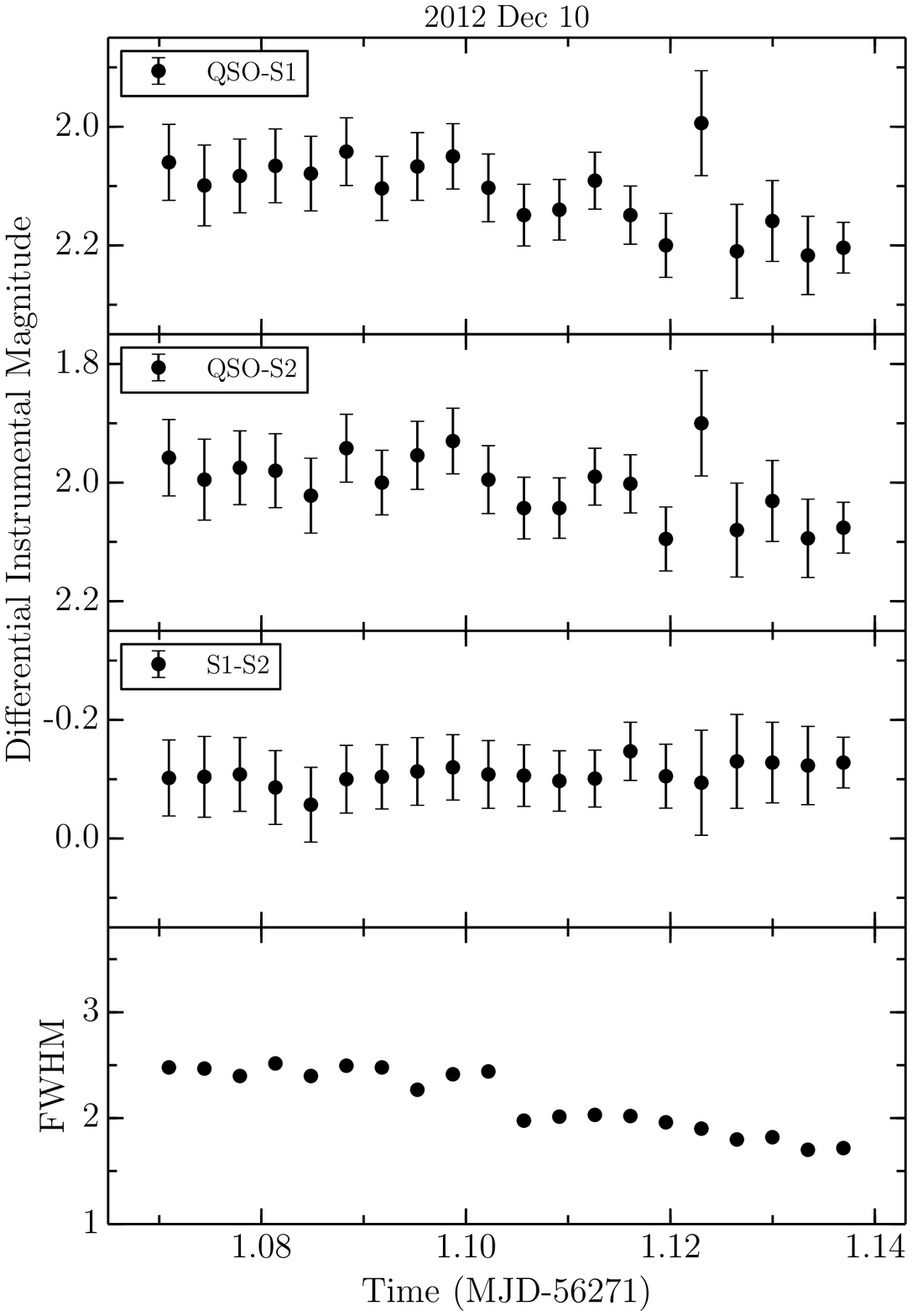}
}
\hbox{
\includegraphics[width=5.5cm]{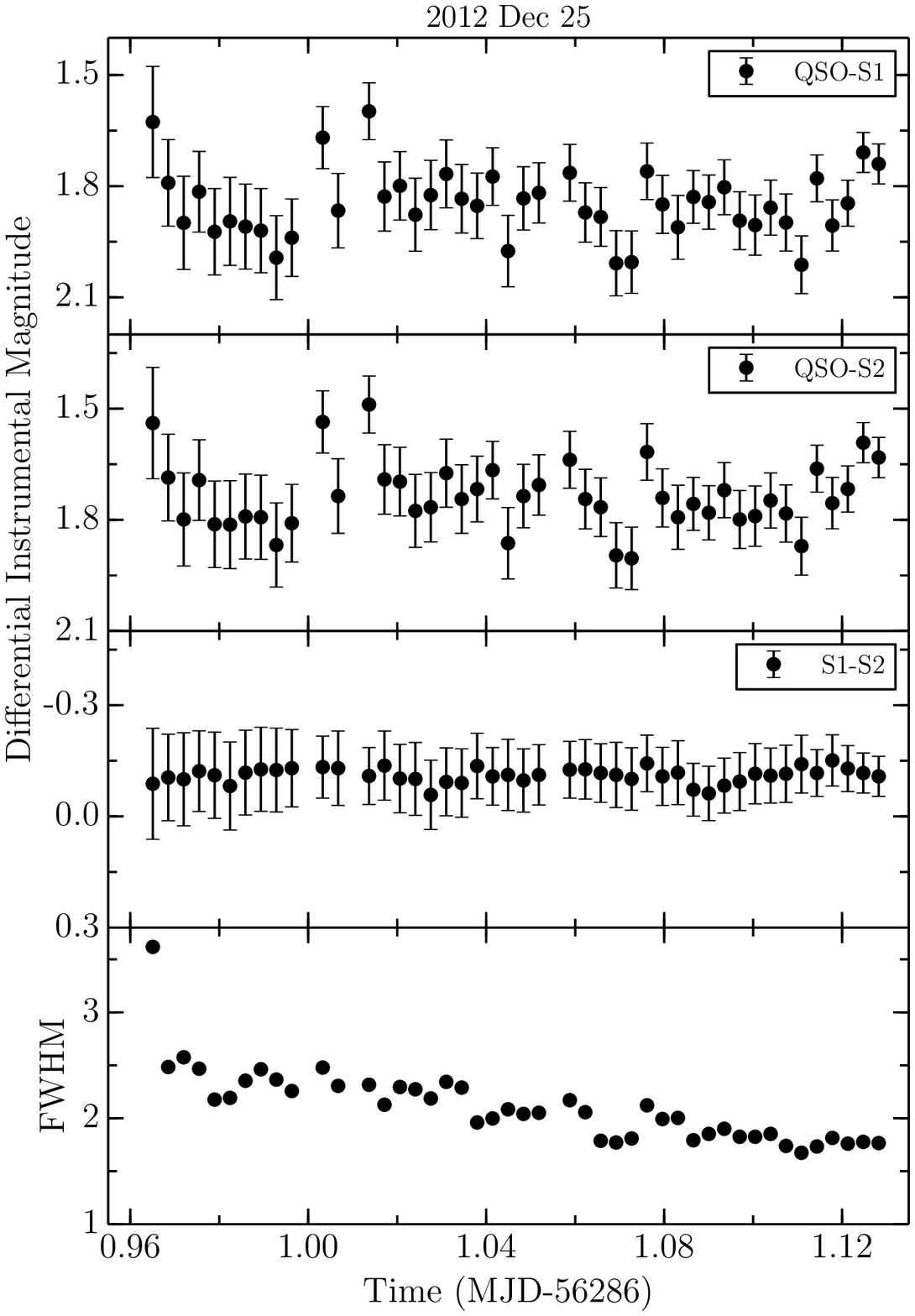}
\includegraphics[width=5.5cm]{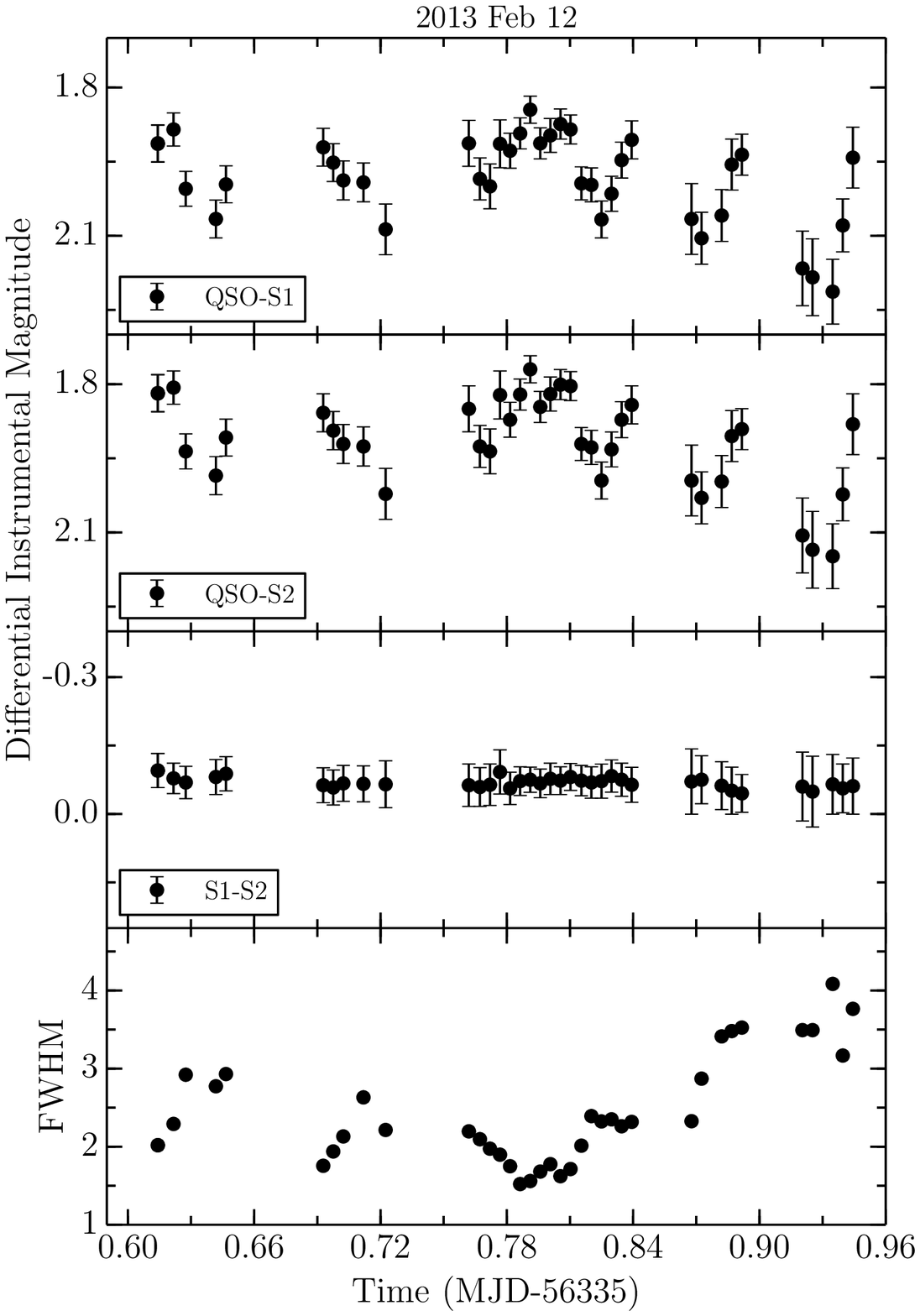}
\includegraphics[width=5.5cm]{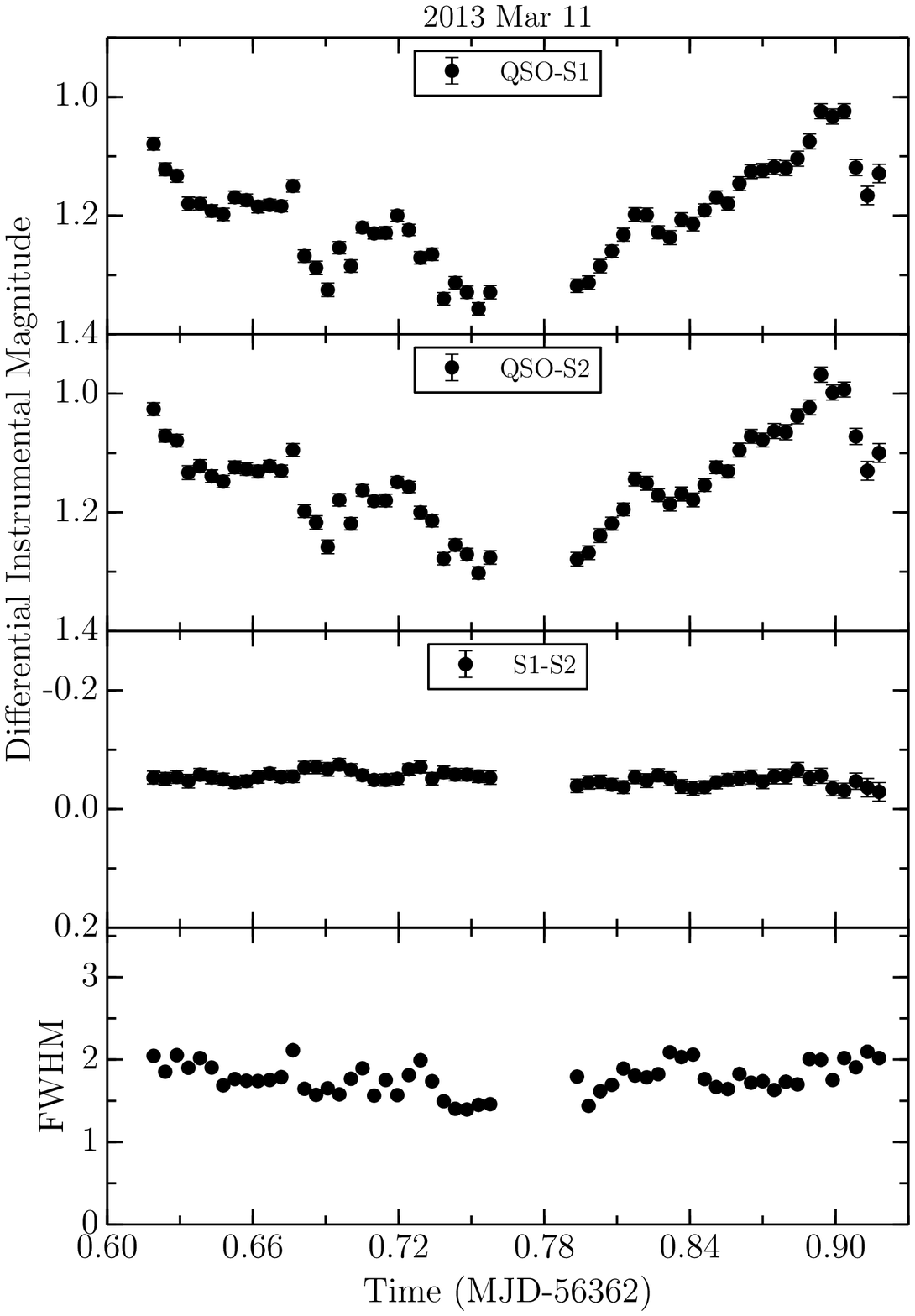}
}
\caption{Intra-night differential light curves of SBS 0846+513. On the bottom panel of each plot is given the variations of the FWHM of the stellar images during the night. QSO refers to the source SBS 0846+513, while S1 and S2 denote the two comparison stars selected to generate the differential light curves.\label{fig_inov}} 
\end{figure}

We observed SBS 0846+513 on a total of six nights between 2012 November to 
2013 March, using the 2m Himalayan Chandra Telescope and the 1.3 m telescope
at Devasthal. The differential light curves of the NLSy1 galaxy relative to 
each of the two comparison stars and the comparison stars themselves are
shown in Figure~\ref{fig_inov}.  
The source is considered as variable only when it shows correlated variations 
both in time and amplitude relative to the selected pair of comparison stars. 
To quantify INOV, we use two statistical tests, {\it C}-statistics and 
{\it F}-statistics. The {\it C} parameter is defined 
as follows
\begin{equation}
C = \frac{\sigma_{\rm T}}{\sigma_{\rm CS}}
\end{equation}
 where $\sigma_{\rm T}$ and $\sigma_{\rm CS}$ are the standard deviations of 
the source and the comparison star differential light curves, respectively. The 
source is considered to have shown INOV if {\it C} $\geq$2.576, which 
corresponds to 99\% confidence level \citep{1997AJ....114..565J}.  On the other 
hand, the method of {\it F}-statistics takes into account the ratio of two 
variances given as 
\begin{equation}
F = \frac{\sigma^{2}_{T}}{\sigma^{2}_{\rm CS}}
\end{equation}
 where $\sigma^{2}_{T}$ is the variance of source-comparison star differential 
light curves and $\sigma^{2}_{\rm CS}$ is the variance of the comparison 
star-star differential light curves. We then compare the derived {\it F} 
values with critical {\it F} value, {\it F$^{\alpha}_{\nu}$}, where $\alpha$ 
is the significance level\footnote{ A significance level of $\alpha$ = 0.01 
corresponds to a confidence level $>$ 99\%.} and $\nu$ (= N$_{p}-$ 1) is the 
degree of freedom for the light curve. We assume the source to be variable 
only if both the computed {\it F} values, with respect to the differential 
light curves of the source to each of the two comparison stars, are greater 
than the critical F{\it } value. The results of {\it F} and {\it C} statistics
are given in Table~\ref{tab_inov}. It should be noted that {\it C}-statistics might 
be a more appropriate quantity to check for the presence of variability, 
especially when the comparison star light curves are not steady. Further, 
we have not found any correlation between the variability pattern of full 
width at half maximum (FWHM) of the point source with that observed in the 
source light curves. Therefore we conclude that the observed INOV are true 
flux variations 
of SBS 0846+513. As can be seen in Figure~\ref{fig_inov} and also in Table~\ref{tab_inov}, the source 
has shown significant flux variations on all the nights. This suggests a duty 
cycle of 100\%, using both {\it C} and {\it F} statistics. The duty cycle is defined as the ratio of the time over which the object shows flux variations to the total observing time. It is calculated as follows
\begin{equation}
DC  = 100\frac{\sum_{i=1}^n N_i(1/\Delta t_i)}{\sum_{i=1}^n (1/\Delta t_i)} {\rm ~\%},
\end{equation}
where $\Delta t_{i}$ = $\Delta t_{i,{\rm obs}}(1 + z)^{-1}$ is the duration of the monitoring session of a source on the {\it i}th night, corrected for the redshift $z$. $N_i$ is equal to 1 if INOV is detected and otherwise 0. Considering the INOV patterns exhibited by other \nlsy~galaxies, studied in our earlier work \citep[][]{2013MNRAS.428.2450P,2014ApJ...789..143P}, the overall duty cycle for \nlsy~galaxies is found to be 55\% and 81\%, according to {\it C} and {\it F} statistics, respectively. Such high amplitude ($\psi >$3\%), high duty cycle ($\sim$70\%) INOV are generally seen in the BL Lac objects \citep[e.g.,][]{2004JApA...25....1S}. This supports the idea that the INOV characteristics of \nlsy~galaxies are similar to blazars.

An interesting phenomenon is the observation of INOV behavior of a source 
during the \gm-ray flaring period. It has been shown by \citet{2014ApJ...789..143P}, based on the results of another \nlsy~galaxy 1H 0323+342, that the chances of detecting large amplitude INOV are higher during the \gm-ray 
flaring period. Coincidentally, one of the INOV observations reported here 
(2013 March 11) was taken during one of the $\gamma$-ray flares of 
SBS 0846+513 (period F2). During this night, the brightness of the source first 
decreased by $\sim$0.3 magnitude and then increased by a similar amount within $\sim$7 hrs.
(see Figure\ref{fig_inov}). As can be seen in this plot, on top of the large flare, many 
small amplitude but fast variations are also visible. Such fast 
INOV features are also seen in the  $\gamma$-NLSy1 galaxies 1H 0323+342 and 
PMN J0948+0022 \citep[][]{2013MNRAS.428.2450P,2014ApJ...789..143P}. This hints for the existence of high incidence of INOV 
during $\gamma$-ray flaring activity, and consequently argue for a jet 
based origin for the observed INOV.

\section{Conclusions}\label{sec_con}
In this work, we present a multi-wavelength study of the \gm-ray emitting 
NLSy1 galaxy SBS 0846+513. The INOV observations reported here are new
and the  observation on any particular night span more than five hours.
We summarize our findings below.
\begin{enumerate}
\item The multi-wavelength light curves of the source show multiple episodes 
of \gm-ray flaring activities which are also reflected in lower frequency 
observations. However, the existence of close correlation between \gm-ray and
X-ray wavelengths could not be statistically confirmed due to the sparseness of 
the data.
\item A statistically significant curvature is noticed in the seven years 
averaged LAT spectrum of SBS 0846+513, a feature generally seen in 
the \gm-ray spectra of powerful FSRQs.
\item The broadband SEDs of SBS 0846+513 show the typical double hump 
structure, similar to blazars, with a steep optical and a relatively 
hard \gm-ray spectrum.
\item The modeling of the SEDs with a one zone leptonic emission model 
indicates the optical-UV spectrum to be dominated by synchrotron emission, 
whereas, the X-ray spectra are well explained by SSC process. The \gm-ray 
regions of the SEDs are explained by IC scattering of BLR photons. It is 
found that during both the flaring activity states, the emission region was 
located at the outer edge of the BLR, while during the low activity state it was well within the BLR.
\item Significant INOV is noticed on all the six nights of ground based 
optical observations. Interestingly, a large amplitude INOV is seen 
during the \gm-ray flaring period F2.
\end{enumerate}

\acknowledgments
We thank the referee for constructive comments that helped in improving the manuscript. This research has made use of the data obtained from HEASARC provided by the 
NASA's Goddard Space Flight Center. Part of this work is based on archival 
data, software or on-line services provided by the ASI Science Data 
Center (ASDC). This research has made use of the XRT Data Analysis Software 
(XRTDAS) developed under the responsibility of the ASDC, Italy. Use of 
{\it Hydra} cluster at Indian Institute of Astrophysics is acknowledged.

\bibliographystyle{apj}
\bibliography{Master}

\newpage
\begin{table*}
\centering
\caption{Positions and apparent magnitudes of the comparison stars from the USNO catalog}\label{tab_stars}
\begin{tabular}{lcccc}
\hline
Star & RA (J2000) & Dec. (J2000) & R\\
 &  (h m s) & (d m s) & (mag)\\
\hline
S1  & 08:49:38.49 & +51:07:55.52 & 16.90\\
S2  & 08:49:41.34 & +51:10:57.97 & 16.90\\
S3  & 08:50:16.73 & +51:09:23.36 & 17.40\\
\hline
\end{tabular}
\end{table*}

\begin{table*}
\begin{center}
{
\small
\caption{Seven years averaged LAT data analysis of \nlsy~galaxy SBS 0846+513. Column information are as follows: (1) model used for fitting (PL: power law, LP: log parabola); (2) 0.1$-$300 GeV flux in units of 10$^{-8}$ \phflux; (3) power law photon index or log parabolic photon index at pivot energy obtained from fitting the log parabola model; (4) curvature index; (5) TS; and (6) $\gamma$-ray luminosity in units of 10$^{46}$ \lum.\label{tab:gamma}}
\begin{tabular}{cccccc}
\hline\hline
 Model & F$_{0.1-300}$ GeV &  $\Gamma_{0.1-300}$/$\alpha$ & $\beta$ & TS & L$_{\gamma}$\\
(1)    & (2)               & (3)       & (4)     & (5)&  (6)        \\
\hline
PL & 3.79 $\pm$ 0.16 & 2.23 $\pm$ 0.03 & ---             & 2634.82 & 4.25\\
LP & 3.18 $\pm$ 0.19 & 1.96 $\pm$ 0.06 & 0.11 $\pm$ 0.02 & 2648.15 & 4.30\\
\hline
\end{tabular}
}
\end{center}
\end{table*}

\begin{table}
\begin{center}
{
\footnotesize
\caption{Summary of SED generation analysis.\label{tab:sed_flux}}
\begin{tabular}{ccccccc}
\hline\hline
 & & & {\it Fermi}-LAT &  & & \\
 Activity state & Period$^{1}$ & Flux$^{2}$ & Photon Index$^{3}$ & Test Statistic$^{4}$ & N$_{\rm pred}^{5}$ & \\
 \hline
 F1 & 56000$-$56200  &9.04~$\pm$~0.61 & 2.11~$\pm$~0.04 & 1085.60 & 1056.01  & \\
 Q  & 56200$-$56380  &1.53~$\pm$~0.56 & 2.56~$\pm$~0.26 & 17.32  & 150.97  & \\
 F2 & 56380$-$56500  &9.92~$\pm$~0.84 & 2.10~$\pm$~0.05 & 732.15 & 658.01  & \\
 \hline
 & &  & {\it Swift}-XRT  & &  & \\
 Activity state & Exp.$^{6}$ & Photon index$^{7}$ & Flux$^{8}$ & Normalization$^{9}$& Stat.$^{10}$ &  \\
\hline
 F1 & 4.75 & 1.45~$\pm$~0.28 & 1.02$^{+0.32}_{-0.21}$ & 1.15~$\pm$~0.26 & 81.42/77 & \\
 Q  & 19.14& 1.61~$\pm$~0.23 & 0.39$^{+0.10}_{-0.07}$ & 0.52~$\pm$~0.01 & 105.14/131& \\
 F2 & 7.63 & 1.34~$\pm$~0.16 & 1.68$^{+0.31}_{-0.25}$ & 1.69~$\pm$~0.23 & 142.28/177 & \\
 \hline
 & & & {\it Swift}-UVOT & & &  \\
 Activity state & V$^{11}$& B$^{11}$& U$^{11}$& W1$^{11}$& M2$^{11}$& W2$^{11}$ \\
 \hline
 F1 & 3.67~$\pm$~0.20& 2.71~$\pm$~0.13& 1.82~$\pm$~0.09& 0.95~$\pm$~0.06& 1.02~$\pm$~0.06& 0.79~$\pm$~0.04 \\
 Q  & 0.62~$\pm$~0.10& 0.39~$\pm$~0.06& 0.18~$\pm$~0.06& 0.16~$\pm$~0.02& 0.16~$\pm$~0.02& 0.14~$\pm$~0.01 \\
 F2 & 7.57~$\pm$~0.26& 5.81~$\pm$~0.23& 3.82~$\pm$~0.18& 2.35~$\pm$~0.15& 2.80~$\pm$~0.17& 1.86~$\pm$~0.10 \\
 \hline
\end{tabular}
}
\end{center}
$^{1}$Time interval considered for SED modeling, in MJD.\\
$^{2}$Integrated $\gamma$-ray flux in 0.1$-$300 GeV energy range in units of 10$^{-8}$ \phflux.\\
$^{3}$Photon index calculated from $\gamma$-ray analysis.\\
$^{4}$Significance of detection using likelihood analysis.\\
$^{5}$Number of predicted photons during the time period under consideration.\\
$^{6}$Net exposure in kiloseconds.\\
$^{7}$Photon index derived from X-ray analysis.\\
$^{8}$Observed flux in units of 10$^{-12}$ \ergflux, in 0.3$-$10 keV energy band.\\
$^{9}$Normalization at 1 keV in 10$^{-4}$ \phflux~keV$^{-1}$.\\
$^{10}$Statistical parameters: C-stat./dof.\\
$^{11}$Average flux in {\it Swift} UVOT bands in units of 10$^{-12}$ \ergflux.\\
\end{table}

\begin{table}
\begin{center}
\caption{Summary of the parameters used/derived from the SED modeling. Viewing angle is assumed as 3$^{\circ}$ and the characteristic temperature of the dusty torus as 900 K. For a disk luminosity of 1 $\times$ 10$^{44}$ \lum~and black hole mass of 5.5 $\times$ 10$^7$ $M_{\odot}$, the size of the BLR is 0.01 parsec (1947 $R_{\rm Sch}$).\label{tab_sed_param}}
\begin{tabular}{llccc}
\hline
\hline
Parameter                     & Symbol   &  F1          & Q            & F2    \\
\hline
Particle spectral index before break energy & $p$                            & 1.8          & 1.9          & 1.8      \\
Particle spectral index after break energy & $q$                            & 7.0          & 7.2          & 7.0      \\
Magnetic field in Gauss & $B$                            & 3.5          & 4.2          & 3.9      \\
Particle energy density in erg cm$^{-3}$ & $U'_{\rm e}$                   & 0.40         & 0.70         & 0.42     \\
Bulk Lorentz factor & $\Gamma$                       & 13           & 7            & 13       \\
Break Lorentz factor of electrons & $\gamma'_{\rm brk}$            & 1088         & 1047         & 1278      \\
Maximum Lorentz factor of electrons & $\gamma'_{\rm max}$            & 3e4          & 3e4          & 3e4      \\
Distance of the emission region in parsec ($R_{\rm Sch}$) & $R_{\rm diss}$                 & 0.011 (2100) & 0.009 (1700) & 0.011 (2100)\\
\hline
Jet power in electrons in log scale & $P_{\rm e}$                    & 43.87        & 43.39        & 43.89  \\
Magnetic jet power in log scale & $P_{\rm B}$                    & 43.95        & 43.38        & 44.05  \\
Radiative jet power in log scale & $P_{\rm r}$                    & 44.50        & 43.81        & 44.66  \\
Jet power in protons in log scale & $P_{\rm p}$                    & 46.04        & 45.69        & 46.05  \\
\hline
\end{tabular}
\end{center}
\end{table}

\begin{table}[h!]
\caption{Log of INOV observations of SBS 0846+513. Column information are as follows: [1] observation date; [2] INOV amplitude in percent; [3]  and [4] {\it C}-values computed for differential light curves relative to the steadiest pair of comparison stars; [5] variability status according to {\it C}-statistics, V: variable, NV: non-variable; [6] and [7] Values of {\it F} parameter for the differential light curves relative to the two comparison stars; and [8] variability status as per {\it F}-statistics.\label{tab_inov}} 
\begin{center}
\begin{tabular}{ccrrcrrc}
\hline \hline
Date     & $\psi$ & {\it C1}  & {\it C2} & Status  & {\it F1}  & {\it F2}  &  Status \\
yyyy mmm dd & (percent) &           &          &         &           &           &         \\
~[1]     & [2]    & [3]       & [4]      & [5]     & [6]       & [7]       & [8]    \\
\hline
2012 Nov 20 & 11.99 & 2.62 & 2.93 & V & 6.85   & 8.60   & V  \\
2012 Dec 9   & 24.37 & 6.98 & 6.88 & V & 48.78 & 47.33 & V  \\
2012 Dec 10 & 22.15 & 3.32 & 2.92 & V & 11.00 & 8.53   & V  \\
2012 Dec 25 & 41.30 & 4.55 & 4.51 & V & 20.67 & 20.32 & V  \\
2013 Feb 12 & 36.76 & 7.56 & 7.91 & V & 57.15 & 62.50 & V  \\
2013 Mar 11 & 33.27 & 7.91 & 7.64 & V & 62.53 & 58.38 & V  \\
 \hline
\end{tabular}
\end{center}
\end{table}

\newpage

\end{document}